\documentclass[conference,10pt,twocolumn,a4paper]{IEEEtran}

\usepackage{cite}
\usepackage[T1]{fontenc}
\usepackage{graphicx}
\usepackage{amssymb}
\usepackage{amsmath}
\usepackage{paralist}
\usepackage{subfigure}
\usepackage{booktabs} 
\usepackage{algorithm}
\usepackage{algorithmic}
\usepackage{amsthm}
\usepackage{multirow}
\usepackage{microtype}
\usepackage{tablefootnote}
\usepackage{hyperref}
\usepackage{framed}
\usepackage{balance}

\usepackage{amssymb}
\usepackage{amsfonts}
\usepackage{mathrsfs}
\usepackage{xspace}
\usepackage{bm}
\usepackage{upgreek}

\newcommand{\safemath}[2]{\newcommand{#1}{\ensuremath{#2}\xspace}}

\safemath{\bma}{\mathbf{a}}
\safemath{\bmb}{\mathbf{b}}
\safemath{\bmc}{\mathbf{c}}
\safemath{\bmd}{\mathbf{d}}
\safemath{\bme}{\mathbf{e}}
\safemath{\bmf}{\mathbf{f}}
\safemath{\bmg}{\mathbf{g}}
\safemath{\bmh}{\mathbf{h}}
\safemath{\bmi}{\mathbf{i}}
\safemath{\bmj}{\mathbf{j}}
\safemath{\bmk}{\mathbf{k}}
\safemath{\bml}{\mathbf{l}}
\safemath{\bmm}{\mathbf{m}}
\safemath{\bmn}{\mathbf{n}}
\safemath{\bmo}{\mathbf{o}}
\safemath{\bmp}{\mathbf{p}}
\safemath{\bmq}{\mathbf{q}}
\safemath{\bmr}{\mathbf{r}}
\safemath{\bms}{\mathbf{s}}
\safemath{\bmt}{\mathbf{t}}
\safemath{\bmu}{\mathbf{u}}
\safemath{\bmv}{\mathbf{v}}
\safemath{\bmw}{\mathbf{w}}
\safemath{\bmx}{\mathbf{x}}
\safemath{\bmy}{\mathbf{y}}
\safemath{\bmz}{\mathbf{z}}
\safemath{\bmzero}{\mathbf{0}}
\safemath{\bmone}{\mathbf{1}}

\bmdefine{\biad}{a}
\bmdefine{\bibd}{b}
\bmdefine{\bicd}{c}
\bmdefine{\bidd}{d}
\bmdefine{\bied}{e}
\bmdefine{\bifd}{f}
\bmdefine{\bigd}{g}
\bmdefine{\bihd}{h}
\bmdefine{\biid}{i}
\bmdefine{\bijd}{j}
\bmdefine{\bikd}{k}
\bmdefine{\bild}{l}
\bmdefine{\bimd}{m}
\bmdefine{\bind}{n}
\bmdefine{\biod}{o}
\bmdefine{\bipd}{p}
\bmdefine{\biqd}{q}
\bmdefine{\bird}{r}
\bmdefine{\bisd}{s}
\bmdefine{\bitd}{t}
\bmdefine{\biud}{u}
\bmdefine{\bivd}{v}
\bmdefine{\biwd}{w}
\bmdefine{\bixd}{x}
\bmdefine{\biyd}{y}
\bmdefine{\bizd}{z}

\bmdefine{\bixid}{\xi}
\bmdefine{\bilambdad}{\lambda}
\bmdefine{\bimud}{\mu}
\bmdefine{\bithetad}{\theta}
\bmdefine{\biphid}{\phi}
\bmdefine{\bideltad}{\delta}

\safemath{\bmia}{\biad}
\safemath{\bmib}{\bibd}
\safemath{\bmic}{\bicd}
\safemath{\bmid}{\bidd}
\safemath{\bmie}{\bied}
\safemath{\bmif}{\bifd}
\safemath{\bmig}{\bigd}
\safemath{\bmih}{\bihd}
\safemath{\bmii}{\biid}
\safemath{\bmij}{\bijd}
\safemath{\bmik}{\bikd}
\safemath{\bmil}{\bild}
\safemath{\bmim}{\bimd}
\safemath{\bmin}{\bind}
\safemath{\bmio}{\biod}
\safemath{\bmip}{\bipd}
\safemath{\bmiq}{\biqd}
\safemath{\bmir}{\bird}
\safemath{\bmis}{\bisd}
\safemath{\bmit}{\bitd}
\safemath{\bmiu}{\biud}
\safemath{\bmiv}{\bivd}
\safemath{\bmiw}{\biwd}
\safemath{\bmix}{\bixd}
\safemath{\bmiy}{\biyd}
\safemath{\bmiz}{\bizd}

\safemath{\bmxi}{\bixid}
\safemath{\bmlambda}{\bilambdad}
\safemath{\bmmu}{\bimud}
\safemath{\bmtheta}{\bithetad}
\safemath{\bmphi}{\biphid}
\safemath{\bmdelta}{\bideltad}

\safemath{\bA}{\mathbf{A}}
\safemath{\bB}{\mathbf{B}}
\safemath{\bC}{\mathbf{C}}
\safemath{\bD}{\mathbf{D}}
\safemath{\bE}{\mathbf{E}}
\safemath{\bF}{\mathbf{F}}
\safemath{\bG}{\mathbf{G}}
\safemath{\bH}{\mathbf{H}}
\safemath{\bI}{\mathbf{I}}
\safemath{\bJ}{\mathbf{J}}
\safemath{\bK}{\mathbf{K}}
\safemath{\bL}{\mathbf{L}}
\safemath{\bM}{\mathbf{M}}
\safemath{\bN}{\mathbf{N}}
\safemath{\bO}{\mathbf{O}}
\safemath{\bP}{\mathbf{P}}
\safemath{\bQ}{\mathbf{Q}}
\safemath{\bR}{\mathbf{R}}
\safemath{\bS}{\mathbf{S}}
\safemath{\bT}{\mathbf{T}}
\safemath{\bU}{\mathbf{U}}
\safemath{\bV}{\mathbf{V}}
\safemath{\bW}{\mathbf{W}}
\safemath{\bX}{\mathbf{X}}
\safemath{\bY}{\mathbf{Y}}
\safemath{\bZ}{\mathbf{Z}}

\safemath{\bZero}{\mathbf{0}}
\safemath{\bOne}{\mathbf{1}}
\safemath{\bDelta}{\mathbf{\Delta}}
\safemath{\bLambda}{\mathbf{\UpLambda}}
\safemath{\bPhi}{\mathbf{\Upphi}}
\safemath{\bSigma}{\mathbf{\Upsigma}}
\safemath{\bOmega}{\mathbf{\Upomega}}
\safemath{\bTheta}{\mathbf{\Uptheta}}

\bmdefine{\biAd}{A}
\bmdefine{\biBd}{B}
\bmdefine{\biCd}{C}
\bmdefine{\biDd}{D}
\bmdefine{\biEd}{E}
\bmdefine{\biFd}{F}
\bmdefine{\biGd}{G}
\bmdefine{\biHd}{H}
\bmdefine{\biId}{I}
\bmdefine{\biJd}{J}
\bmdefine{\biKd}{K}
\bmdefine{\biLd}{L}
\bmdefine{\biMd}{M}
\bmdefine{\biOd}{N}
\bmdefine{\biPd}{O}
\bmdefine{\biQd}{P}
\bmdefine{\biRd}{R}
\bmdefine{\biSd}{S}
\bmdefine{\biTd}{T}
\bmdefine{\biUd}{U}
\bmdefine{\biVd}{V}
\bmdefine{\biWd}{W}
\bmdefine{\biXd}{X}
\bmdefine{\biYd}{Y}
\bmdefine{\biZd}{Z}

\bmdefine{\biDelta}{\Delta}
\bmdefine{\biLambda}{\Lambda}
\bmdefine{\biPhi}{\Phi}
\bmdefine{\biSigma}{\Sigma}
\bmdefine{\biOmega}{\Omega}
\bmdefine{\biTheta}{\Theta}

\safemath{\bimA}{\biAd}
\safemath{\bimB}{\biBd}
\safemath{\bimC}{\biCd}
\safemath{\bimD}{\biDd}
\safemath{\bimE}{\biEd}
\safemath{\bimF}{\biFd}
\safemath{\bimG}{\biGd}
\safemath{\bimH}{\biHd}
\safemath{\bimI}{\biId}
\safemath{\bimJ}{\biJd}
\safemath{\bimK}{\biKd}
\safemath{\bimL}{\biLd}
\safemath{\bimM}{\biMd}
\safemath{\bimN}{\biNd}
\safemath{\bimO}{\biOd}
\safemath{\bimP}{\biPd}
\safemath{\bimQ}{\biQd}
\safemath{\bimR}{\biRd}
\safemath{\bimS}{\biSd}
\safemath{\bimT}{\biTd}
\safemath{\bimU}{\biUd}
\safemath{\bimV}{\biVd}
\safemath{\bimW}{\biWd}
\safemath{\bimX}{\biXd}
\safemath{\bimY}{\biYd}
\safemath{\bimZ}{\biZd}

\safemath{\bimDelta}{\biDelta}
\safemath{\bimLambda}{\biLambda}
\safemath{\bimPhi}{\biPhi}
\safemath{\bimSigma}{\biSigma}
\safemath{\bimOmega}{\biOmega}
\safemath{\bimTheta}{\biTheta}

\safemath{\setA}{\mathcal{A}}
\safemath{\setB}{\mathcal{B}}
\safemath{\setC}{\mathcal{C}}
\safemath{\setD}{\mathcal{D}}
\safemath{\setE}{\mathcal{E}}
\safemath{\setF}{\mathcal{F}}
\safemath{\setG}{\mathcal{G}}
\safemath{\setH}{\mathcal{H}}
\safemath{\setI}{\mathcal{I}}
\safemath{\setJ}{\mathcal{J}}
\safemath{\setK}{\mathcal{K}}
\safemath{\setL}{\mathcal{L}}
\safemath{\setM}{\mathcal{M}}
\safemath{\setN}{\mathcal{N}}
\safemath{\setO}{\mathcal{O}}
\safemath{\setP}{\mathcal{P}}
\safemath{\setQ}{\mathcal{Q}}
\safemath{\setR}{\mathcal{R}}
\safemath{\setS}{\mathcal{S}}
\safemath{\setT}{\mathcal{T}}
\safemath{\setU}{\mathcal{U}}
\safemath{\setV}{\mathcal{V}}
\safemath{\setW}{\mathcal{W}}
\safemath{\setX}{\mathcal{X}}
\safemath{\setY}{\mathcal{Y}}
\safemath{\setZ}{\mathcal{Z}}
\safemath{\emptySet}{\varnothing}

\safemath{\colA}{\mathscr{A}}
\safemath{\colB}{\mathscr{B}}
\safemath{\colC}{\mathscr{C}}
\safemath{\colD}{\mathscr{D}}
\safemath{\colE}{\mathscr{E}}
\safemath{\colF}{\mathscr{F}}
\safemath{\colG}{\mathscr{G}}
\safemath{\colH}{\mathscr{H}}
\safemath{\colI}{\mathscr{I}}
\safemath{\colJ}{\mathscr{J}}
\safemath{\colK}{\mathscr{K}}
\safemath{\colL}{\mathscr{L}}
\safemath{\colM}{\mathscr{M}}
\safemath{\colN}{\mathscr{N}}
\safemath{\colO}{\mathscr{O}}
\safemath{\colP}{\mathscr{P}}
\safemath{\colQ}{\mathscr{Q}}
\safemath{\colR}{\mathscr{R}}
\safemath{\colS}{\mathscr{S}}
\safemath{\colT}{\mathscr{T}}
\safemath{\colU}{\mathscr{U}}
\safemath{\colV}{\mathscr{V}}
\safemath{\colW}{\mathscr{W}}
\safemath{\colX}{\mathscr{X}}
\safemath{\colY}{\mathscr{Y}}
\safemath{\colZ}{\mathscr{Z}}

\safemath{\opA}{\mathbb{A}}
\safemath{\opB}{\mathbb{B}}
\safemath{\opC}{\mathbb{C}}
\safemath{\opD}{\mathbb{D}}
\safemath{\opE}{\mathbb{E}}
\safemath{\opF}{\mathbb{F}}
\safemath{\opG}{\mathbb{G}}
\safemath{\opH}{\mathbb{H}}
\safemath{\opI}{\mathbb{I}}
\safemath{\opJ}{\mathbb{J}}
\safemath{\opK}{\mathbb{K}}
\safemath{\opL}{\mathbb{L}}
\safemath{\opM}{\mathbb{M}}
\safemath{\opN}{\mathbb{N}}
\safemath{\opO}{\mathbb{O}}
\safemath{\opP}{\mathbb{P}}
\safemath{\opQ}{\mathbb{Q}}
\safemath{\opR}{\mathbb{R}}
\safemath{\opS}{\mathbb{S}}
\safemath{\opT}{\mathbb{T}}
\safemath{\opU}{\mathbb{U}}
\safemath{\opV}{\mathbb{V}}
\safemath{\opW}{\mathbb{W}}
\safemath{\opX}{\mathbb{X}}
\safemath{\opY}{\mathbb{Y}}
\safemath{\opZ}{\mathbb{Z}}
\safemath{\opZero}{\mathbb{O}}
\safemath{\identityop}{\opI}

\safemath{\veca}{\bma}
\safemath{\vecb}{\bmb}
\safemath{\vecc}{\bmc}
\safemath{\vecd}{\bmd}
\safemath{\vece}{\bme}
\safemath{\vecf}{\bmf}
\safemath{\vecg}{\bmg}
\safemath{\vech}{\bmh}
\safemath{\veci}{\bmi}
\safemath{\vecj}{\bmj}
\safemath{\veck}{\bmk}
\safemath{\vecl}{\bml}
\safemath{\vecm}{\bmm}
\safemath{\vecn}{\bmn}
\safemath{\veco}{\bmo}
\safemath{\vecp}{\bmp}
\safemath{\vecq}{\bmq}
\safemath{\vecr}{\bmr}
\safemath{\vecs}{\bms}
\safemath{\vect}{\bmt}
\safemath{\vecu}{\bmu}
\safemath{\vecv}{\bmv}
\safemath{\vecw}{\bmw}
\safemath{\vecx}{\bmx}
\safemath{\vecy}{\bmy}
\safemath{\vecz}{\bmz}

\safemath{\veczero}{\bmzero}
\safemath{\vecone}{\bmone}
\safemath{\vecxi}{\bmxi}
\safemath{\veclambda}{\bmlambda}
\safemath{\vecmu}{\bmmu}
\safemath{\vectheta}{\bmtheta}
\safemath{\vecphi}{\bmphi}
\safemath{\vecdelta}{\bmdelta}

\safemath{\matA}{\bA}
\safemath{\matB}{\bB}
\safemath{\matC}{\bC}
\safemath{\matD}{\bD}
\safemath{\matE}{\bE}
\safemath{\matF}{\bF}
\safemath{\matG}{\bG}
\safemath{\matH}{\bH}
\safemath{\matI}{\bI}
\safemath{\matJ}{\bJ}
\safemath{\matK}{\bK}
\safemath{\matL}{\bL}
\safemath{\matM}{\bM}
\safemath{\matN}{\bN}
\safemath{\matO}{\bO}
\safemath{\matP}{\bP}
\safemath{\matQ}{\bQ}
\safemath{\matR}{\bR}
\safemath{\matS}{\bS}
\safemath{\matT}{\bT}
\safemath{\matU}{\bU}
\safemath{\matV}{\bV}
\safemath{\matW}{\bW}
\safemath{\matX}{\bX}
\safemath{\matY}{\bY}
\safemath{\matZ}{\bZ}
\safemath{\matzero}{\bmzero}

\safemath{\matDelta}{\bDelta}
\safemath{\matLambda}{\bLambda}
\safemath{\matPhi}{\bPhi}
\safemath{\matSigma}{\bSigma}
\safemath{\matOmega}{\bOmega}
\safemath{\matTheta}{\bTheta}

\safemath{\matidentity}{\matI}
\safemath{\matone}{\matO}

\safemath{\rnda}{A}
\safemath{\rndb}{B}
\safemath{\rndc}{C}
\safemath{\rndd}{D}
\safemath{\rnde}{E}
\safemath{\rndf}{F}
\safemath{\rndg}{G}
\safemath{\rndh}{H}
\safemath{\rndi}{I}
\safemath{\rndj}{J}
\safemath{\rndk}{K}
\safemath{\rndl}{L}
\safemath{\rndm}{M}
\safemath{\rndn}{N}
\safemath{\rndo}{O}
\safemath{\rndp}{P}
\safemath{\rndq}{Q}
\safemath{\rndr}{R}
\safemath{\rnds}{S}
\safemath{\rndt}{T}
\safemath{\rndu}{U}
\safemath{\rndv}{V}
\safemath{\rndw}{W}
\safemath{\rndx}{X}
\safemath{\rndy}{Y}
\safemath{\rndz}{Z}

\safemath{\rveca}{\bimA}
\safemath{\rvecb}{\bimB}
\safemath{\rvecc}{\bimC}
\safemath{\rvecd}{\bimD}
\safemath{\rvece}{\bimE}
\safemath{\rvecf}{\bimF}
\safemath{\rvecg}{\bimG}
\safemath{\rvech}{\bimH}
\safemath{\rveci}{\bimI}
\safemath{\rvecj}{\bimJ}
\safemath{\rveck}{\bimK}
\safemath{\rvecl}{\bimL}
\safemath{\rvecm}{\bimM}
\safemath{\rvecn}{\bimN}
\safemath{\rveco}{\bomO}
\safemath{\rvecp}{\bimP}
\safemath{\rvecq}{\bimQ}
\safemath{\rvecr}{\bimR}
\safemath{\rvecs}{\bimS}
\safemath{\rvect}{\bimT}
\safemath{\rvecu}{\bimU}
\safemath{\rvecv}{\bimV}
\safemath{\rvecw}{\bimW}
\safemath{\rvecx}{\bimX}
\safemath{\rvecy}{\bimY}
\safemath{\rvecz}{\bimZ}

\safemath{\rvecxi}{\bmxi}
\safemath{\rveclambda}{\bmlambda}
\safemath{\rvecmu}{\bmmu}
\safemath{\rvectheta}{\bmtheta}
\safemath{\rvecphi}{\bmphi}

\safemath{\rmatA}{\bimA}
\safemath{\rmatB}{\bimB}
\safemath{\rmatC}{\bimC}
\safemath{\rmatD}{\bimD}
\safemath{\rmatE}{\bimE}
\safemath{\rmatF}{\bimF}
\safemath{\rmatG}{\bimG}
\safemath{\rmatH}{\bimH}
\safemath{\rmatI}{\bimI}
\safemath{\rmatJ}{\bimJ}
\safemath{\rmatK}{\bimK}
\safemath{\rmatL}{\bimL}
\safemath{\rmatM}{\bimM}
\safemath{\rmatN}{\bimN}
\safemath{\rmatO}{\bimO}
\safemath{\rmatP}{\bimP}
\safemath{\rmatQ}{\bimQ}
\safemath{\rmatR}{\bimR}
\safemath{\rmatS}{\bimS}
\safemath{\rmatT}{\bimT}
\safemath{\rmatU}{\bimU}
\safemath{\rmatV}{\bimV}
\safemath{\rmatW}{\bimW}
\safemath{\rmatX}{\bimX}
\safemath{\rmatY}{\bimY}
\safemath{\rmatZ}{\bimZ}

\safemath{\rmatDelta}{\bimDelta}
\safemath{\rmatLambda}{\bimLambda}
\safemath{\rmatPhi}{\bimPhi}
\safemath{\rmatSigma}{\bimSigma}
\safemath{\rmatOmega}{\bimOmega}
\safemath{\rmatTheta}{\bimTheta}

\usepackage{amssymb}
\usepackage{amsfonts}
\usepackage{mathrsfs}
\usepackage{xspace}
\usepackage{bm}
\usepackage{fancyref}
\usepackage{textcomp}

\usepackage{multirow}
\usepackage{stmaryrd}

\newenvironment{textbmatrix}{	\setlength{\arraycolsep}{2.5pt}%
								\big[\begin{matrix}}{\end{matrix}\big]%
								\raisebox{0.08ex}{\vphantom{M}}}

\def\be{\begin{equation}}
\def\ee{\end{equation}}
\def\een{\nonumber \end{equation}}
\def\mat{\begin{bmatrix}}
\def\emat{\end{bmatrix}}
\def\btm{\begin{textbmatrix}}
\def\etm{\end{textbmatrix}}

\def\ba#1\ea{\begin{align}#1\end{align}}
\def\bas#1\eas{\begin{align*}#1\end{align*}}
\def\bs#1\es{\begin{split}#1\end{split}}
\def\bg#1\eg{\begin{gather}#1\end{gather}}
\def\bml#1\eml{\begin{multline}#1\end{multline}}
\def\bi#1\ei{\begin{itemize}#1\end{itemize}}

\newcommand{\lefto}{\mathopen{}\left}

\DeclareMathOperator*{\argmin}{arg\;min}		
\DeclareMathOperator{\Exop}{\opE}			
\DeclareMathOperator{\Varop}{\opV\!\mathrm{ar}} 

\newcommand{\Ex}[2]{\ensuremath{\Exop_{#1}\lefto[#2\right]}} 	
\newcommand{\abs}[1]{\lefto\lvert#1\right\rvert}		

\newcommand{\vecnorm}[1]{\lefto\lVert#1\right\rVert}		

\safemath{\dirac}{\delta}					
\safemath{\krond}{\dirac}					

\safemath{\upto}{\uparrow}
\safemath{\downto}{\downarrow}
\safemath{\iu}{j}							
\safemath{\ev}{\lambda}						
\safemath{\hilseqspace}{l^{2}}				
\newcommand{\banachfunspace}[1]{\setL^{#1}}	
\safemath{\hilfunspace}{\banachfunspace{2}}	

\safemath{\SNR}{\textit{SNR}} 				
\safemath{\PAR}{\textit{PAR}} 				
\safemath{\No}{N_0}							
\safemath{\Es}{E_s}							
\safemath{\Eb}{E_b}							
\safemath{\EbNo}{\frac{\Eb}{\No}}
\safemath{\EsNo}{\frac{\Es}{\No}}

\DeclareMathOperator{\CHop}{\ensuremath{\opH}} 
\safemath{\tvir}{\rndh_{\CHop}}				
\safemath{\tvtf}{\rndl_{\CHop}}				
\safemath{\spf}{\rnds_{\CHop}}				
\safemath{\bff}{H_{\CHop}}					

\safemath{\ircf}{r_{h}}						
\safemath{\tftvcf}{r_{s}}					
\safemath{\tfcf}{r_{l}}						
\safemath{\bfcf}{r_{H}}						

\safemath{\tcorr}{c_h}						
\safemath{\scf}{c_{s}}						
\safemath{\tfcorr}{c_{l}}					
\safemath{\fcorr}{c_{H}}						

\safemath{\mi}{I}							
\safemath{\capacity}{C}						

\safemath{\normal}{\mathcal{N}}			
\safemath{\jpg}{\mathcal{CN}}			
\safemath{\mchain}{\leftrightarrow}		

\safemath{\dB}{\,\mathrm{dB}}
\safemath{\dBm}{\,\mathrm{dBm}}
\safemath{\Hz}{\,\mathrm{Hz}}
\safemath{\kHz}{\,\mathrm{kHz}}
\safemath{\MHz}{\,\mathrm{MHz}}
\safemath{\GHz}{\,\mathrm{GHz}}
\safemath{\s}{\,\mathrm{s}}
\safemath{\ms}{\,\mathrm{ms}}
\safemath{\mus}{\,\mathrm{\text{\textmu}s}}
\safemath{\ns}{\,\mathrm{ns}}
\safemath{\ps}{\,\mathrm{ps}}
\safemath{\meter}{\,\mathrm{m}}
\safemath{\mm}{\,\mathrm{mm}}
\safemath{\cm}{\,\mathrm{cm}}
\safemath{\m}{\,\mathrm{m}}
\safemath{\W}{\,\mathrm{W}}
\safemath{\mW}{\, \mathrm{mW}}
\safemath{\J}{\,\mathrm{J}}
\safemath{\K}{\,\mathrm{K}}
\safemath{\bit}{\,\mathrm{bit}}
\safemath{\nat}{\,\mathrm{nat}}

\safemath{\define}{\triangleq}			

\safemath{\equivalent}{\sim}
\safemath{\distas}{\sim}					
\safemath{\sdiff}{\Delta}				

\safemath{\reals}{\mathbb{R}}
\safemath{\positivereals}{\reals_{+}}
\safemath{\integers}{\mathbb{Z}}
\safemath{\posint}{\integers_{+}}
\safemath{\naturals}{\mathbb{N}}
\safemath{\posnaturals}{\naturals_{+}}
\safemath{\complexset}{\mathbb{C}}
\safemath{\rationals}{\mathbb{Q}}

\newcommand*{\fancyrefapplabelprefix}{app}		
\newcommand*{\fancyrefthmlabelprefix}{thm}		
\newcommand*{\fancyreflemlabelprefix}{lem}		
\newcommand*{\fancyrefcorlabelprefix}{cor}		
\newcommand*{\fancyrefdeflabelprefix}{def}		
\newcommand*{\fancyrefproplabelprefix}{prop}		
\newcommand*{\fancyrefexmpllabelprefix}{exmpl}
\newcommand*{\fancyrefalglabelprefix}{alg}		
\newcommand*{\fancyreftbllabelprefix}{tbl}		

\frefformat{vario}{\fancyrefseclabelprefix}{Section~#1}
\frefformat{vario}{\fancyrefthmlabelprefix}{Theorem~#1}
\frefformat{vario}{\fancyreftbllabelprefix}{Table~#1}
\frefformat{vario}{\fancyreflemlabelprefix}{Lemma~#1}
\frefformat{vario}{\fancyrefcorlabelprefix}{Corollary~#1}
\frefformat{vario}{\fancyrefdeflabelprefix}{Definition~#1}
\frefformat{vario}{\fancyreffiglabelprefix}{Fig.~#1}
\frefformat{vario}{\fancyrefapplabelprefix}{Appendix~#1}
\frefformat{vario}{\fancyrefeqlabelprefix}{(#1)}
\frefformat{vario}{\fancyrefproplabelprefix}{Proposition~#1}
\frefformat{vario}{\fancyrefexmpllabelprefix}{Example~#1}
\frefformat{vario}{\fancyrefalglabelprefix}{Algorithm~#1}

 \newtheorem{defi}{Definition}

\safemath{\dictab}{[\,\dicta\,\,\dictb\,]}

\safemath{\ysig}{\bmy}
\safemath{\ysighat}{\hat{\ysig}}
\safemath{\ysigdim}{M}
\safemath{\xsig}{\bmx}
\safemath{\xsigdim}{N}
\safemath{\nx}{n_x}
\safemath{\zsig}{\bmz}
\safemath{\zsigdim}{\ysigdim}
\safemath{\rsig}{\bmr}
\safemath{\Adict}{\bA}
\safemath{\Adicttilde}{\widetilde{\Adict}}
\safemath{\Adictdim}{\outputdim\times\xsigdim}
\safemath{\avec}{\bma}
\safemath{\avectilde}{\tilde{\avec}}
\safemath{\Bdict}{\bB}
\safemath{\Bdicttilde}{\widetilde{\Bdict}}
\safemath{\Cdict}{\bC}
\safemath{\cvec}{\bmc}
\safemath{\Ddict}{\bD}
\safemath{\Ddictdim}{\ysigdim\times\xsigdim}
\safemath{\dvec}{\bmd}
\safemath{\Ddicttilde}{\widetilde{\bD}}
\safemath{\Bonb}{\bB}
\safemath{\bvec}{\bmb}
\safemath{\Bonbdim}{\ysigdim\times\ysigdim}
\safemath{\noise}{\bmn}
\safemath{\noisedim}{\ysigim}
\safemath{\err}{\bme}
\safemath{\errdim}{\ysigdim}
\safemath{\errset}{\setE}
\safemath{\nerr}{n_e}
\safemath{\delop}{\bP_\errset}
\safemath{\delopc}{\bP_{{\errset}^c}}

\safemath{\cplxi}{\imath}
\safemath{\cplxj}{\jmath}

\safemath{\dict}{\matD}
\safemath{\inputdim}{N}		
\safemath{\outputdim}{M}		
\safemath{\sparsity}{S}	
\safemath{\inputdimA}{{N_a}}	
\safemath{\inputdimB}{{N_b}}	
\safemath{\elemA}{{n_a}}	
\safemath{\elemB}{{n_b}}	
\safemath{\resA}{\matR_a}	
\safemath{\resB}{\matR_b}	
\safemath{\subD}{\matS} 
\safemath{\subA}{\matS_a} 
\safemath{\subB}{\matS_b} 
\safemath{\dicta}{\matA} 	
\safemath{\dictb}{\matB} 	
\safemath{\hollowS}{H}
\safemath{\hollowA}{H_a}
\safemath{\hollowB}{H_b}
\safemath{\cross}{Z}
\safemath{\coh}{\mu_d}			
\safemath{\coha}{\mu_a}			
\safemath{\cohb}{\mu_b}			
\safemath{\mubs}{\nu}	
\safemath{\cohm}{\mu_m} 
\safemath{\dictset}{\setD}	
\safemath{\dictsetp}{\dictset(\coh,\coha,\cohb)}	
\safemath{\dictsetgen}{\dictset_\text{gen}}
\safemath{\dictsetgenp}{\dictsetgen(\coh)}
\safemath{\dictsetonb}{\dictset_\text{onb}}
\safemath{\dictsetonbp}{\dictsetonb(\coh)}

\safemath{\leftside}{U}
\safemath{\rightsideA}{R_a}
\safemath{\rightsideB}{R_b}

\safemath{\indexS}{\setI_S} 

\safemath{\na}{n_a}			
\safemath{\nb}{n_b}			
\safemath{\coeffa}{p_i}	
\safemath{\coeffb}{q_j}	
\safemath{\seta}{\setP}		
\safemath{\setb}{\setQ}     
\safemath{\setw}{\setW}	
\safemath{\setz}{\setZ}	
\safemath{\cola}{\veca}		
\safemath{\colb}{\vecb}		
\safemath{\cold}{\vecd}		
\safemath{\inputvec}{\vecx} 	
\safemath{\error}{\vece}	
\safemath{\noiseout}{\vecz} 	
\safemath{\inputvecel}{x}
\safemath{\inputveca}{\vecx_a}
\safemath{\inputvecb}{\vecx_b}
\safemath{\outputvec}{\vecy}	
\safemath{\lambdamin}{\lambda_{\mathrm{min}}}

\safemath{\elltwo}{\ell_2}
\safemath{\ellone}{\ell_1}
\safemath{\ellzero}{\ell_0}
\safemath{\ellinf}{\ell_\infty}
\safemath{\ellinftilde}{\ell_{\widetilde\infty}}
\safemath{\licard}{Z(\coh,\coha,\cohb)}
\safemath{\xsol}{\hat{x}}
\safemath{\xbord}{x_b}		
\safemath{\xstat}{x_s}		
\safemath{\xstatLone}{\tilde{x}_s}
\safemath{\order}{\mathcal{O}} 
\safemath{\scales}{\Theta} 
\safemath{\ones}{\mathbf{1}} 
\safemath{\zeroes}{\mathbf{0}} 
\safemath{\thlone}{\kappa(\coh,\cohb)} 
\safemath{\constoneA}{\delta} 
\safemath{\constoneB}{\epsilon} 
\safemath{\nlarge}{L}				   
\safemath{\sumlarge}{S_\nlarge}
\safemath{\maxlarger}{P_\nlarge}	   
\safemath{\Pzero}{\textrm{P0}}	
\safemath{\Pone}{\textrm{P1}}
\safemath{\vecfir}{\vecw}			 
\safemath{\vecsec}{\vecz}
\safemath{\elvecfir}{w}              
\safemath{\elvecsec}{z}				 
\safemath{\nlargefir}{n}
\safemath{\normout}{\gamma}
\safemath{\auxfun}{h}
\safemath{\supp}{\textrm{supp}}

\safemath{\indexa}{\ell}
\safemath{\indexb}{r}
\safemath{\indexc}{i}
\safemath{\indexd}{j}

\safemath{\project}{P}

\usepackage{color}

\usepackage{enumitem}
\setlist[itemize]{leftmargin=*, itemsep=0.3em, topsep=0.3em} 

\allowdisplaybreaks 
\IEEEoverridecommandlockouts

\safemath{\LAMA}{\textrm{LAMA}}
\safemath{\MRT}{\textrm{MRT}}
\safemath{\betamax}{\beta^\text{max}_\setO}
\safemath{\betamaxno}{\beta^\text{max}}
\safemath{\betamin}{\beta^\text{min}_\setO}
\safemath{\betaminno}{\beta^\text{min}}

\safemath{\Nomin}{\No^\textnormal{min}(\beta)}
\safemath{\Nominnobeta}{\No^\text{min}}
\safemath{\Nomax}{\No^\textnormal{max}(\beta)}
\safemath{\Nomaxnobeta}{\No^\textnormal{max}}
\safemath{\EX}{E_\textnormal{x}}
\safemath{\EXP}{\EX^\textnormal{p}}

\safemath{\tmax}{{t_\textnormal{max}}}
\safemath{\MAP}{\textrm{MAP}}
\safemath{\IO}{\textrm{IO}}
\safemath{\JO}{\textrm{JO}}
\safemath{\Nopost}{N_{0}^\textnormal{post}}
\safemath{\MT}{U}
\safemath{\MR}{B}
\safemath{\Tran}{\textnormal{T}}
\safemath{\Herm}{\textnormal{H}}
\safemath{\row}{\textnormal{r}}
\safemath{\col}{\textnormal{c}}

\safemath{\NT}{N_\textnormal{T}}
\safemath{\DSNR}{\delta \textnormal{SNR}}
\safemath{\betaMOR}{\beta^{\star}}

\usepackage{array}
\newcolumntype{C}[1]{>{\centering\let\newline\\\arraybackslash\hspace{0pt}}m{#1}}

\begin{document}
	
\title{VLSI Design of a Nonparametric Equalizer \\ for Massive MU-MIMO
}

\author{
\IEEEauthorblockN{Charles Jeon$^{1}$, Gulnar Mirza$^{1}$, Ramina Ghods$^{1}$, Arian Maleki$^{2}$, and Christoph Studer$^{1}$}\\ \vspace{-0.1cm}
\IEEEauthorblockA{$^\text{1}$Cornell University, Ithaca, NY;  \{cj339,\,gzm3,\,rg548,\,studer\}@cornell.edu; web: \url{http://vip.ece.cornell.edu}} 
\IEEEauthorblockA{$^\text{2}$Columbia University, New York City, NY; {arian@stat.columbia.edu}} \\[-0.65cm]
\thanks{CJ, GM, RG, and CS were supported in part by Xilinx, Inc.~and by the US National Science Foundation (NSF) under grants ECCS-1408006, CCF-1535897, CAREER CCF-1652065, and CNS-1717559.}
\thanks{The authors would like to thank A.\ Burg and R.\ Manohar for discussions on Cannon's algorithm and O. Casta\~neda for his help with the MVU.}
}

\maketitle

\begin{abstract}
Linear minimum mean-square error (L-MMSE) equalization is among the most popular  methods for data detection  in massive multi-user multiple-input multiple-output (MU-MIMO) wireless systems. 
While L-MMSE equalization  enables near-optimal spectral efficiency, accurate knowledge of the signal and noise powers is necessary. Furthermore, corresponding VLSI designs must solve linear systems of equations, which requires high arithmetic precision, exhibits stringent data dependencies, and results in high circuit complexity.
This paper proposes the first VLSI design of the NOnParametric Equalizer (NOPE), which avoids knowledge of the transmit signal and noise powers, provably delivers the performance of  L-MMSE equalization  for massive MU-MIMO systems, and  is resilient to numerous system and hardware impairments due to its parameter-free nature.
Moreover, NOPE avoids computation of a matrix inverse and only requires hardware-friendly matrix-vector multiplications.
To showcase the practical advantages of NOPE, we propose a parallel VLSI architecture   and provide synthesis results in 28\,nm CMOS.
We demonstrate that NOPE performs on par with existing data detectors for massive MU-MIMO that require accurate knowledge of the signal and noise powers.
%
\end{abstract}


\section{Introduction}
It is widely believed that massive multi-user multiple-input multiple-output (MU-MIMO) will be a core technology for fifth-generation~(5G) wireless systems.
Massive MU-MIMO relies on base-station (BS) architectures with hundreds of antenna elements and radio-frequency (RF) chains that serve tens of user equipments (UEs) in the same time-frequency resource.
While this emerging technology enables unprecedented spectral efficiency by means of fine-grained beamforming~\cite{LETM2014,ABCHLAZ2014}, it also poses significant practical implementation challenges.

\subsection{The Case for Nonparametric Equalization}

Data detection in the uplink (UEs transmit data to the BS) is among the most critical tasks from a spectral efficiency and hardware complexity perspective~\cite{WYWDCS2014}. 
While optimal MIMO data detection is known to be  NP-hard~\cite{V1998}, it has been shown in~\cite{HBD11,WYWDCS2014} that linear minimum mean-square error (L-MMSE) equalization enables near-optimal performance in massive MU-MIMO systems. 
However, L-MMSE equalization requires accurate knowledge of the signal and noise powers~\cite{GJMAS17}.
Furthermore, corresponding hardware designs must solve linear systems of equations, which requires high arithmetic precision and suffers from stringent data dependencies---both of these aspects result in relatively high circuit complexity~\cite{studer2011asic,CJDCGS2017,PRLE2017,TCZ2016,PLZYW2017}. 
In addition, practical massive MU-MIMO BS designs will most likely rely on inexpensive RF circuitry which suffers from numerous impairments, including amplifier nonlinearities, phase noise, and quantization artifacts~\cite{Studer_Tx_OFDM,bjornson2014massive}. The presence of non-ideal hardware necessitates the design of new equalization algorithms that are resilient to real-world hardware imperfections. 

Recently,  a novel algorithm called \emph{NonParametric Equalizer} (NOPE, for short) was
proposed in~\cite{GJMAS17}. NOPE does not require knowledge of the signal and noise powers while provably achieving the performance of L-MMSE equalization in massive MU-MIMO systems. 
NOPE combines approximate message passing (AMP)~\cite{donoho2009message} with Stein's unbiased risk estimator (SURE)~\cite{stein1981estimation} and mismatched data detection~\cite{JMS2016}, which renders this algorithm resilient to numerous hardware impairments while being computationally efficient: NOPE only requires matrix-vector products and avoids a computation of costly matrix inverses or matrix decompositions, which are typically required by L-MMSE equalizer algorithms. 
Despite all these advantages, NOPE has been designed only for idealistic channel models and has not yet been integrated in hardware.

\subsection{Contributions}
In this paper, we generalize NOPE to practical channels and provide, to the best of our knowledge, its first VLSI design. 
Our contributions are summarized as follows:
\begin{itemize}
\item We propose a set of algorithm-level modifications that enable NOPE to operate on more realistic MU channels.
\item We develop a VLSI architecture that relies on Cannon's algorithm~\cite{CannonThesis} to achieve high throughput at low area.
\item We show reference VLSI synthesis results in 28\,nm CMOS for a 64 BS antenna, 16 UE massive MU-MIMO system.
\item We compare NOPE to existing massive MU-MIMO equalizers requiring knowledge of the signal and noise powers.
\end{itemize}
Our results demonstrate that massive MU-MIMO has the unique potential to design parameter-free algorithms, such as NOPE, that perform on par with solutions that require accurate knowledge of critical system and model parameters.

\subsection{Notation}
Lowercase and uppercase boldface letters designate column vectors and matrices, respectively. 
The transpose and conjugate transpose of a matrix $\bA$ are denoted by $\bA^\text{T}$ and $\bA^\Herm$, respectively; the entry on the $k$th row and $\ell$th column of $\bA$ is $A_{k,\ell}$.
The $M\times M$ identity matrix is denoted by~$\bI_M$ and the $M\times N$ all-zeros matrix by $\mathbf{0}_{M\times N}$. 
The $k$th entry of a vector $\veca$ is~$a_k$. 
We define the Hadamard product as~$\circ$.
For an $N$-dimensional vector~$\bma$, we define $\left\langle \bma \right\rangle = \frac{1}{N}\sum_{k=1}^N a_k$. 
The probability density function (PDF) of a circularly-symmetric complex-valued Gaussian random vector with covariance matrix~$\bK$ is denoted by $\setC\setN(\bZero,\bK)$. Expectation and variance with respect to the random vector~$\bma$ is denoted by~$\Exop_\bma\!\left[\cdot\right]$ 
and $\Varop_\bma\!\left[ \cdot\right] $, respectively. 

%
%
%
%
%
%
%
%
%

%

\section{A Primer on L-MMSE Equalization}\label{sec:L-equalization}
We now introduce the system model and review the basics of L-MMSE equalization.  
We then discuss NOPE.

\subsection{System Model}
We consider the input-output relation  $\bmy=\bH \bmx+\bmn$ to model a massive MU-MIMO uplink system operating in a frequency-flat channel~\cite{WYWDCS2014}.
The vector $\vecy\in\complexset^\MR$ contains the received signals at the BS; $\MR$ denotes the number of BS antennas; the matrix $\bH\in\complexset^{\MR\times \MT}$ represents the uplink MIMO channel; $\MT$ denotes the number of UEs; the transmit signal vector is $\vecx\in\complexset^\MT$; and the vector $\bmn\in\complexset^\MR$ models receive noise, which has i.i.d.\ circularly-symmetric complex Gaussian entries with variance~$\No$ per entry. 
Throughout this paper, we assume that the transmit signal vector~$\bmx$ has i.i.d. entries so that $p(\vecx)=\prod_{u=1}^{\MT}p(x_u)$, where~$p(\cdot)$ models the signal prior (e.g., a 16-QAM constellation) with zero mean and signal variance $\EX=\Ex{}{|x_u|^2}$, $u=1,\ldots,U$. We require  the following definitions. 
\begin{defi}
The \emph{antenna ratio} is defined as $\beta=\MT/\MR$.
\end{defi}
\begin{defi}
The \emph{large-antenna limit} is defined by fixing the antenna ratio $\beta$ and letting $\MT\to\infty$.
\end{defi}
\begin{defi}
The channel matrix $\bH$ is said to have \emph{uniform channel gains} if the entries are i.i.d.\ circularly-symmetric complex Gaussian with variance $1/\MR$ per complex entry.
\end{defi}

\subsection{Basics of L-MMSE Equalization}\label{sec:L-MMSE}
L-MMSE equalization is among the most popular methods to compute an estimate for $\bmx$ from $\bmy$ and from knowledge of the channel matrix~$\bH$, and enjoys widespread use for data detection in MIMO systems~\cite{madhow1994mmse,kumar2009asymptotic,studer2011asic,hoydis2011massive}. 
The relatively low computational complexity (except for the inversion of a potentially large matrix) and acceptable performance render this method a feasible alternative to more complicated data detection algorithms.
Moreover, it has been shown in~\cite{HBD11,WYWDCS2014} that L-MMSE equalization enables (often significantly) higher achievable rates than zero-forcing (ZF) or maximum ratio combining (MRC)-based equalizers in massive MU-MIMO systems.
However, to enable near-optimal spectral efficiency via L-MMSE equalization, accurate knowledge of the signal and noise powers is required; see, e.g., 
\cite{TH1999,GJMAS17}. 

Mathematically, the goal of L-MMSE equalization is to compute a linear estimate $\hat\bmx=\bW\bmy$ from the receive vector $\bmy$ that minimizes the $\textit{MSE} = \Ex{\,\bmx,\bmn}{\|\hat\bmx-\bmx\|^2}$ using knowledge of the channel matrix~$\bH$ as well as the signal and noise powers, $\EX$ and $\No$, respectively. 
For a circularly-symmetric complex-valued transmit signal~$\bmx$, the equalization matrix $\bW\in\complexset^{\MT\times \MR}$ is given by $\bW = (\bH^\Herm \bH+ \frac{\No }{\EX}\bI_U)^{-1} \bH^\Herm$. 
If the signal~$\bmx$ is zero-mean and real-valued (e.g., for BPSK signals), then the optimal linear estimator for the real part $\bmx_\text{Re}$ is given by
\begin{align*}
\hat\bmx_\text{Re} = \textstyle  \!\left(\bH^\text{T}_\text{Re} \bH_\text{Re}+\bH^\text{T}_\text{Im} \bH_\text{Im}+\frac{\No }{2\EX}\bI_U\!\right)^{\!-1}\!\!(\bH^\text{T}_\text{Re}\bmy_\text{Re} + \bH^\text{T}_\text{Im}\bmy_\text{Im}),
\end{align*}
where $\bH_\text{Re}$, $\bH_\text{Im}$,  $\bmy_\text{Re}$, and $\bmy_\text{Im}$ are the real and imaginary parts of $\bH$ and $\bmy$, respectively; the imaginary part of the estimate is $\hat\bmx_\text{Im}=\bZero_{U\times1}$.
Clearly, L-MMSE equalization relies on knowledge of the quantities $\rho={\No }/{\EX}$ or $\rho={\No }/{(2\EX)}$, which requires (i) means to detect whether the  transmit signals are real- or complex-valued and (ii) an accurate estimate of~$\rho$ that is commonly acquired in a dedicated training phase~\cite{perels2007frame}. 
%


\subsection{L-MMSE Equalization via AMP}\label{sec:MMSE-AMP}
As shown in~\cite{JMS2016}, L-MMSE equalization can be implemented using the mismatched complex Bayesian AMP (mcB-AMP) framework. 
By assuming a  mismatched Gaussian signal prior distribution $\tilde{p}(\vecx)=\prod_{i=u}^{U}\tilde{p}(x_u)$ with $\tilde{p}(x_u) \sim \setC\setN(0,\EX)$ instead of the true signal prior $X_0\sim p(x_0)$ (e.g., two Dirac delta functions concentrated at $-1$ and $+1$ for BPSK), one can design the following \emph{parametric} L-MMSE algorithm: 

\newtheorem{alg}{Algorithm}
%
%
\begin{alg}[L-MMSE-AMP~\cite{JMS2016}] Initialize $t=1$, $x^1_u =0$, $u=1,\ldots,\MT$, and $\bmr^1 = \bmy - \bH\bmx^1$. Then, for every iteration $t=1,\ldots,\tmax$ compute the output $\bmz^t$ via the following steps:\label{alg:MMSE-AMP}
   \begin{align}
   \nonumber
   \tilde\sigma^2_{t} &= \textstyle \frac{1}{\MR}\vecnorm{\bmr^{t}}_2^2\\
   \label{eq:tau_opt}
   \tau^{t} &= \textstyle \argmin_{\tau\geq0} \,\Psi(\tilde\sigma_t^2,\tau)\\
   \nonumber
   \bmz^t&=\bmx^{t} + \bH^\Herm \bmr^{t}\\
   \label{eq:xhat_LMMSE}
   \bmx^{t+1} &= \mathsf{F}^\text{mm}\!\left(\bmz^t,\tau^t\right)\\
   \label{eq:residual}
   \bmr^{t+1} &= \bmy - \bH\bmx^{t+1} + \beta \bmr^t \!\left\langle\mathsf{F'}^\text{mm}(\bmz^t,\tau^t)
   \right\rangle\!.
   \end{align}
Here, the posterior mean function $\mathsf{F}^\text{mm}(x_u,\tau)=\frac{\EX}{\EX+\tau}x_u$ and $\mathsf{F'}^\text{mm}(x_u,\tau)=\frac{\EX}{\EX+\tau}$  operate element-wise on vectors. We furthermore need the MSE function:
$\textstyle \Psi(\tilde\sigma_t^2,\tau)=\frac{\tau^2 \EX+\tilde\sigma_t^2\EX^2}{(\EX+\tau)^2}$.
\end{alg}
%
%
%

%
Interestingly, the estimate $\hat\bmz = \lim_{t\to\infty }\bmz^{t}$ computed by L-MMSE-AMP exhibits the same MSE as that of the L-MMSE equalizer  in the large-antenna limit and for matrices~$\bH$ with uniform channel gains~\cite{JMS2016}.
While this is an asymptotic equivalence, reference~\cite{GJMAS17} has shown that the error-rate performance of L-MMSE-AMP is virtually indistinguishable from an   L-MMSE equalizer in practical (finite-dimensional) massive MU-MIMO systems for a small number of iterations~$t_\text{max}$ (ten or fewer). 
Clearly, \fref{alg:MMSE-AMP} mainly relies on matrix-vector multiplications, which enables parallel hardware designs. However, the exact knowledge of $\EX$ is still required.

\section{NOnParametric Equalizer: NOPE}\label{sec:NOPE}
We now summarize the necessary steps to free \fref{alg:MMSE-AMP} from  knowledge of the signal 
power, leading to NOPE. 
We then propose a generalization of the algorithm that makes it suitable for more realistic MIMO system scenarios.

\subsection{The NOPE Algorithm} \label{sec:NOPE1}
To develop NOPE, we wish to automatically tune the signal power~$\EX$ and the parameter $\tau^t$. 
To this end, we first introduce the parameter $\gamma^t={\EX}/{\tau^t}$ and reparametrize the functions $\mathsf{F}^\text{mm}(x_u,\gamma^t)=\frac{\gamma^t}{\gamma^t+1}x_u$ and $\mathsf{F'}^\text{mm}(x_u,\gamma^t)=\frac{\gamma^t}{\gamma^t+1}$ in \fref{alg:MMSE-AMP}. Now, only a single parameter must be tuned per iteration, i.e., $\gamma^t$.
Interestingly, \cite[Thm.~3]{JMS2016} shows that optimal parameter tuning is achieved by tuning each parameter~$\gamma^t$
by minimizing \fref{eq:tau_opt} separately at iteration $t$ starting from $t=1$ to $\tmax$.
Hence, the remaining piece is to replace the MSE function~$\Psi$ with a function that does not depend on the true signal prior $p(x_0)$.
As shown  in~\cite{GJMAS17}, one can use Stein's unbiased risk estimate (SURE)~\cite{stein1981estimation} to extract an estimate  of the  MSE function~$\Psi$ as 
\begin{align}
   \hat{\Psi}(\tilde\sigma_t^2,\gamma^t) = 
   \textstyle
   \, \tilde\sigma_t^2\frac{\gamma^t-1}{\gamma^t+1}+\frac{\|\bmz^t\|_2^2}{\MT(\gamma^t+1)^2}.\label{eq:Psi_hat}
\end{align}
Since the minimum of $\hat{\Psi}$ is given by $\gamma_\text{min}^{t}\!=\!{\|\bmz^t\|_2^2}/(\MT\tilde{\sigma}_t^2)-1$ we can replace the tuning stage in \fref{eq:tau_opt} by $\gamma_\text{min}^{t}$, which leads to NOPE.
As proven in~\cite[Cor.~6]{GJMAS17}, NOPE achieves the performance of an L-MMSE equalizer in the large antenna limit given  that $\bH$ has uniform channel gains and for $t\to\infty$.

\subsection{Robust Version of NOPE}\label{NOPE2}
NOPE and \fref{alg:MMSE-AMP} require the matrix~$\bH$ to have uniform channel gains. However, in practice each UE typically has a different large-scale fading gain (e.g., affected by the distance to the BS), resulting in channel matrices~$\bH$ whose columns have different scale. 
We now show how NOPE can be made robust to such channels. 
As in~\cite{GJMAS17}, one can rewrite the channel matrix as $\bH=\widetilde{\bH}\bD$, where each element of~$\widetilde{\bH}$ is distributed as $\setC\setN(0,1/\MR)$ and $\bD$ is diagonal containing the $u$th UE's individual large-scale fading gain $d_u$.
%
For this model, one must estimate the gain of the $u$th UE  by
$\sum_{b=1}^\MR \abs{H_{b,u}}^2 = {d}_u^2\sum_{b=1}^\MR \vert{\tilde H_{b,u}}\vert^2$, which converges to $d_u^2$ in the large-antenna limit.
%
Thus, $\bD$ is estimated with a diagonal matrix $\widehat{\bD}$, where the $u$th diagonal element is given by 
 $\hat{d}_u$.
To enable NOPE to support nonuniform channel gains, we 
modify the posterior mean function in \fref{eq:xhat_LMMSE} into an element-wise operation~\cite{GJMAS17}
\begin{align}\label{eq:Fmm_l}
\mathsf{F}^\text{mm}_u(z_u^t,\tau^t)= \textstyle \frac{\EX}{\EX+{\tau^t}/{\hat{d}_u^2}}z_u^t.
\end{align}
Furthermore, step \fref{eq:residual} in \fref{alg:MMSE-AMP} must be replaced by 
\begin{align*}
\bmr^{t+1} = \textstyle \bmy - \bH\bmx^{t+1} + \beta \bmr^t \frac{1}{\MT} \sum_{u=1}^\MT \mathsf{F'}^\text{mm}_u(z_u^t,\tau^t)
\end{align*}
to take into account the fact that different functions $\mathsf{F}^\text{mm}_u(z_u^t,\tau^t)$ are used for each UE. 
This generalization also requires new estimates for the parameters~$\EX$ and $\tau^t$ in \fref{eq:Fmm_l}.
As shown in~\cite[Thm~7]{GJMAS17}, both of these parameters can be estimated as follows
\begin{align}
\label{eq:Ex-hat}
\hat{E}_{\text{x}}^t 
= \textstyle \frac{ v_\bmz^t
 - 2 v_\bmr^t}
 {\sum_{u=1}^\MT {\hat{d}_u^2}} \quad \text{and} \quad 
\hat{\tau}^t=\frac{1}{\MR}\|\bmr^t\|^2,
\end{align}
where we introduced shorthand notation $v_\bmz^t =\sum_{u=1}^U \hat{d}_u^2 \abs{z_u^t}^2$ for the weighted-norm of $\bmz^t$ with respect to its large-scale fading gains, and $v_\bmr^t = \beta\vecnorm{\bmr}_2^2/2$ for the residual norm.

The remaining piece of our robust NOPE is to enable \mbox{L-MMSE} data detection for BPSK constellations for which the imaginary part of $\bmx$ is zero. In fact, assuming a circularly-symmetric complex Gaussian prior for BPSK signals is a poor match as the imaginary part is zero.
We generalize NOPE by estimating the signal variance $\EX$ in \fref{eq:Ex-hat} for the real and imaginary parts separately, which enables us to automatically adapt NOPE to the used constellation set. 
To do so, we decompose the weighted-norm of $\bmz^t$ denoted by $v_\bmz^t$, into real and imaginary parts, i.e., $v_\bmz^t = v_{\bmz,\text{Re}}^t + v_{\bmz,\text{Im}}^t$. 
More specifically, we can estimate the necessary variances as 
\begin{align*}
\hat{E}_{\text{x},\text{Re}}^t = \textstyle
\frac{
v_{\bmz,\text{Re}}^t - v_\bmr^t}{\sum_{u=1}^\MT {\hat{d}_u^2}} 
%
\quad \text{and} \quad 
\hat{E}_{\text{x},\text{Im}}^t  =
\frac{
v_{\bmz,\text{Im}}^t - v_\bmr^t
 }{\sum_{u=1}^\MT {\hat{d}_u^2}},
\end{align*}
for which $\hat{E}_{\text{x},\text{Re}}^t+\hat{E}_{\text{x},\text{Im}}^t=\hat{E}_{\text{x}}^t$.
%
%
With all these ingredients, we arrive at the generalized NOPE algorithm in \fref{alg:nope}.

\setlength{\textfloatsep}{10pt}
\setcounter{algorithm}{1}
\begin{algorithm}[tp]
  \caption{Robust version of NOPE} \label{alg:nope}
  \begin{algorithmic}[1]
     \STATE {\bf inputs:} $\bH\in\complexset^{\MR\times \MT}$ and $\bmy\in\complexset^{\MR}$
     \STATE {\bf precompute:} $\hat{d}_u^2 = \sum_{b=1}^\MR \abs{H_{b,u}}^2$, $\hat{d}^{-2}_u=1/\hat{d}_u^2$ for all $u=1,\ldots,\MT$, and $\langle\hat\bmd^2\rangle = \frac{1}{\MT}\sum_{u=1}^\MT\hat{d}_u^2$
     \STATE {\bf initialize:} $t=1$, $\bmx^1=\bm{0}_{\MT\times1}$, and $\langle\bm{\alpha}\rangle = 0$
     \FOR {$t=1,2,\ldots,t_\text{max}$}
     \STATE $\bmr^{t} = \bmy - \bH \bmx^{t} + \frac{\beta}{2} \langle \bm{\alpha} \rangle \bmr^t $
     \hfill (residual update)

     \STATE $v_\bmr^t = \frac{\beta}{2} \vecnorm{\bmr^t}_2^2$ \hfill (norm of $\bmr^t$)
     \STATE $\bmz^t=\bmx^t + \hat{\bmd}^{-2} \circ (\bH^\Herm\bmr^t)$ 
     %

     \STATE $v_{\bmz,\text{Re}}^t = \sum_{u=1}^\MT \hat{d}_u^2 \text{Re}\{z_u^t\}^2$ \hfill (weighted norm of $\text{Re}\{\bmz^t\}$)
     \STATE $v_{\bmz,\text{Im}}^t = \sum_{u=1}^\MT \hat{d}_u^2 \text{Im}\{z_u^t\}^2$ \hfill (weighted norm of $\text{Im}\{\bmz^t\}$)
     \STATE $K^t= (v_\bmr^t \langle \hat\bmd^2\rangle)^{-1}$

     \FOR{$u=1,\ldots,\MT$}
     
     \STATE $\alpha_{u,\text{Re}}^t = (1 + (K^t \hat{d}_u^2 ( v_{\bmz,\text{Re}}^t - v_\bmr^t)) ^{-1})^{-1}$
     \STATE $\alpha_{u,\text{Im}}^t = (1 + (K^t \hat{d}_u^2 ( v_{\bmz,\text{Im}}^t - v_\bmr^t)) ^{-1})^{-1}$
     %
     
     \STATE $x^{t+1}_u = \alpha_{u,\text{Re}}^t \text{Re}\{z_u^t\}+ j \alpha_{u,\text{Im}}^t  \text{Im}\{z_u^t\}$ 

     \STATE $\alpha_u^t = \alpha_{u,\text{Re}}^t + \alpha_{u,\text{Im}}^t$
      \hfill (Onsager constant)

     \STATE $\rho_u^t = (2B/\beta)K^t \langle\hat{\bmd}^2\rangle \hat{d}_u^2 $
      \hfill (post-equalization SNR)

     %
     
     \ENDFOR
     
     \STATE $t=t+1$
     \ENDFOR
     
     \STATE {\bf output:} L-MMSE estimate $\bmz^t$ and   post-equalization SNR values~$\rho_u^t$ for each UE $u=1,\ldots,U$    
  \end{algorithmic}
\end{algorithm}



\subsection{Numerical Results}
\begin{figure}
\centering
\includegraphics[width=0.9\columnwidth]{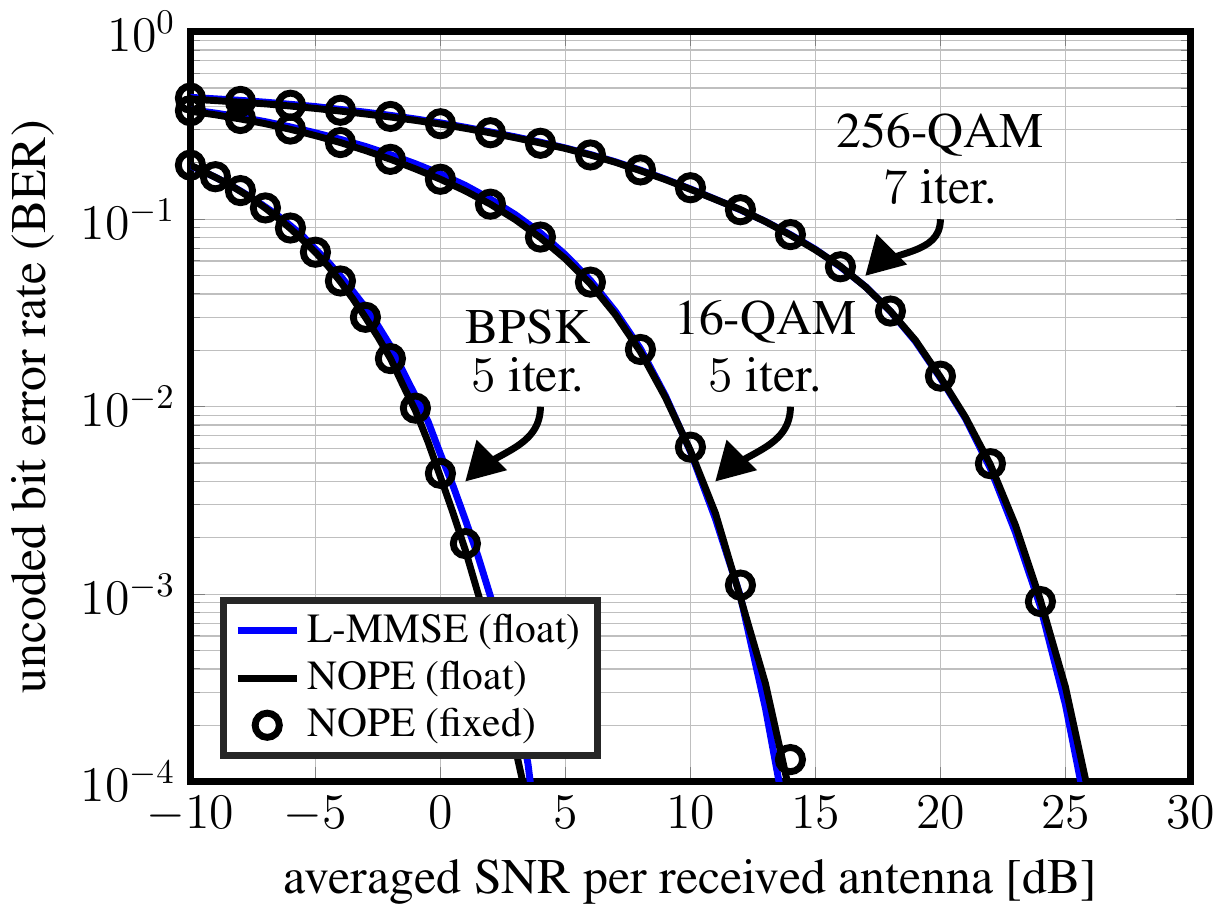}
\vspace{-0.15cm}
\caption
{
Uncoded bit error-rate of NOPE algorithm in a $64\times16$ massive MU-MIMO system with Rayleigh fading channel matrices.
NOPE closely approaches the performance of the L-MMSE estimator that requires exact knowledge of the signal and noise powers.
Furthermore, fixed-point arithmetic in NOPE (shown with circle markers) does not exhibit a significant BER performance loss compared to infinite-precision arithmetic (continuous lines).
}
\label{fig:NOPEC}
\end{figure}
Figure \ref{fig:NOPEC} shows uncoded bit error-rate (BER) simulation results in a $\MR=64$ BS antenna, $\MT=16$ UE massive MU-MIMO system with BPSK, 16-QAM, and 256-QAM. 
We show the performance of exact L-MMSE equalization, as well as the performance of NOPE for both infinite and fixed-precision.
Solid lines correspond to floating-point precision, and circle markers correspond to fixed-point precision simulations of NOPE. 
Evidently, the BER performance of NOPE with $t_\text{max}=5$ iterations ($t_\text{max}=7$ for 256-QAM) is virtually indistinguishable from the exact L-MMSE estimator, which requires accurate knowledge of both the signal and noise powers. Due to its parameter free nature, NOPE is suitable for situations in which the signal and noise powers change rapidly (e.g., due to interference) or if the transmit constellation is unknown and must be estimated prior to data detection.
%


\section{VLSI Architecture and Synthesis Results}

We now propose a very-large scale integration (VLSI) architecture of the NOPE algorithm for a $B=64$ BS antenna, $U=16$ UE massive MU-MIMO system.
We then discuss the most essential optimization steps and finally present  implementation results in a 28\,nm CMOS technology. 

\begin{figure}[t]
\centering
\includegraphics[width = 0.42\textwidth]{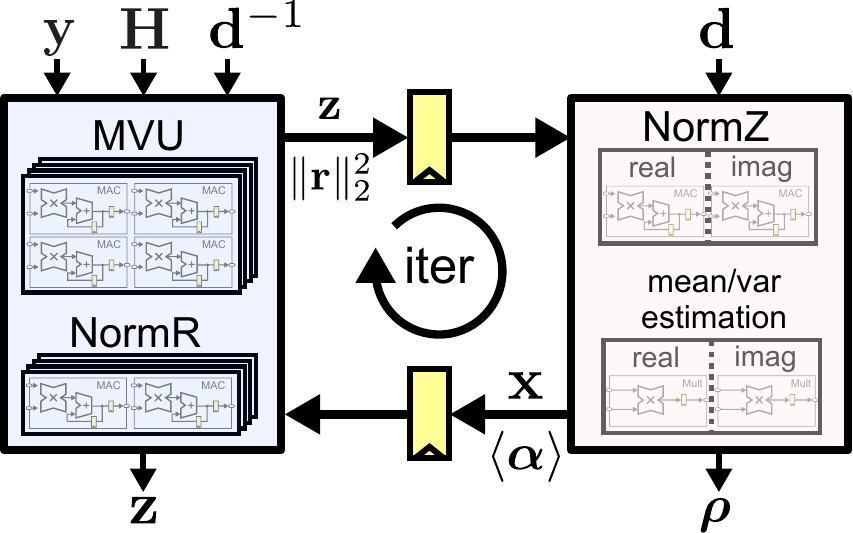}
\vspace{-0.23cm}
\caption{
Each iteration of the NOPE is partitioned into two main units: matrix-vector unit (left) and parameter estimation unit (right).
Both units require an identical number of clock cycles, which allows us to process two independent problems via coarse-grained pipeline interleaving. 
}
\label{fig:toplvl}
\end{figure}

\begin{figure}[t]
\centering
\includegraphics[width = 0.22\textwidth]{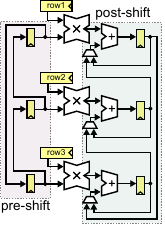}
\vspace{-0.2cm}
\caption{
Illustration of the matrix-vector unit (MVU) that computes computes $\bH\bmx$ and $\bH^\Herm\bmr$ in a $3\times3$ system via Cannon's algorithm.
To compute   $\bH\bmx$, we circularly shift the inputs  (pre-shift); to compute $\bH^\Herm\bmx$, we circularly shift the outputs (post-shift). This approach enables column-wise storage of the entries of $\bH$ without causing access contentions, leading to high throughput. 
}
\label{fig:cannon}
\end{figure}

\subsection{Architecture Overview}
We partition the NOPE iterations into two phases, each executed by a separate unit; see \fref{fig:toplvl} for an architecture overview. The \emph{matrix-vector unit} (MVU) executes the necessary matrix-vector multiplications and the \emph{estimation unit} (EU) implements automatic parameter tuning.

The MVU performs the matrix-vector multiplication required to compute the $16$ dimensional output vector $\bmz$ (line 7 of Alg.~\ref{alg:nope}) and the scalar $\vecnorm{\bmr}_2^2$ (line 5 of Alg. \ref{alg:nope}).
The EU implements the mean and variance estimation to compute the posterior mean~$\bmx$ (line 14 of Alg. \ref{alg:nope}) and Onsager constant $\langle \bm{\alpha}\rangle$ (line 15 of Alg.~\ref{alg:nope}).
In addition, we compute the per-user post-equalization SNR~$\bm\rho$ (line 16 of Alg. \ref{alg:nope}), which is required for log-likelihood ratio (LLR) value calculations to perform soft-output data detection.
Both units carry out their tasks in the same number of clock cycles, which enables us to process two independent equalization problems concurrently in the same architecture by means of coarse-grained pipeline interleaving. 

\subsection{Architecture Details}
We now provide architecture details for the MVU and EU, and briefly discuss the key fixed-point implementation aspects.

\subsubsection{MVU Details}
The MVU computes both $\bH\bmx$ and $\bH^\Herm\bmr$ in a single unified architecture, similarly to the architecture in~\cite{CJDCGS2017}.
We divide the $64\times16$-dimensional channel matrix $\bH$ into four $16\times16$ blocks, each of which are processed using a separate MVU, which we refer to as MVU-$m$, $m\in\{1,2,3,4\}$.
Each $16\times16$ matrix-vector multiplication is carried out using $16$ complex-valued multiply-accumulate (MAC) units; the matrix-vector operation is carried out on a column-by-column basis so that each MAC unit is associated to a row of the matrix.


A straightforward approach to compute $\bH\bmx$ would be to broadcast the $16$-dimensional vector $\bmx$ to all MVUs. To compute $\bH^\Herm\bmr$ within the same architecture, access contentions would arise as one would need to be able to read all entries from the row of $\bH^\Herm$ and sum all partial products. 
To enable highly parallel matrix-vector computation without causing access contentions, we use an architecture as depicted in  \fref{fig:cannon} that performs a variant of Cannon's algorithm \cite{CannonThesis}. 
Let $\bA$ be a $16\times16$ block of $\bH$ where each row $r$ is cyclically shifted by its index. 
To compute $\bA\bmx$, the input $\bmx$ is first loaded into the input shift registers (the pre-shift block). We then circularly shift the entries of this shift register while sequentially calculating the MAC operations with entries of the matrix $\bA$; the outputs are accumulated in the registers at the output of each MAC unit. This effectively implements a column-by-column matrix-vector operation in $16$ clock cycles. 
To compute $\bH^\Herm\bmr$, we load $\bmr$ into the input shift register but no cyclical shifts are carried out. Instead we cyclically exchange the outputs (the post-shift block) while accumulating the results. This effectively implements a row-by-row matrix-vector operation in $16$ clock cycles. 

After the computation of $\bH^\Herm\bmr$, we have to accumulate the results of the four $16\times 16$ blocks. We do this over two additional clock cycles: in cycle 1, MVU-$1$ and MVU-$4$ pass their result to MVU-$2$ and MVU-$3$ for accumulation; in cycle~2, MVU-$2$ passes its result to MVU-$3$ to obtain the final result. 

\subsubsection{EU Details}
The EU computes the posterior mean $\bmx$ and Onsager constant $\langle \bm\alpha\rangle$.
To this end, the EU first computes the $16$-dimensional norm
 of the real and imaginary part of $\bmz$, $v_{\bmz,\text{Re}}$ and $v_{\bmz,\text{Im}}$. We employ two MAC units which compute the real and imaginary parts over 16 clock cycles.
Once $v_{\bmz,\text{Re}}$ and $v_{\bmz,\text{Im}}$ are completed, we compute the so-called denoising parameter $\alpha_{u,\text{Re}}$ (line 12 of Alg. \ref{alg:nope}) and $\alpha_{u,\text{Im}}$ (line 13 of Alg. \ref{alg:nope}) for each $u$-th UE sequentially over 16 clock cycles.
We note that the function $(1+x^{-1})^{-1} = 1- (1+x)^{-1}\in[0,1)$ in lines 12 and 13 of Alg. \ref{alg:nope} is numerically stable so we employ  a single-iteration of the LUT-based Newton-Raphson procedure.

\subsubsection{Fixed-Point Arithmetic}
In order to achieve low hardware complexity and high throughput, our design uses fixed-point arithmetic. 
We first globally scale~$\bH$ so that the real and imaginary entries are less than $1$. We then  quantize each element of $\bH$ to $10$ fraction bits, and $\bmy$ to $6$ integer and $4$ fraction bits.
%
The fixed-point performance of our NOPE design is shown in \fref{fig:NOPEC}. The solid lines correspond to floating-point performance, the markers to the fixed-point performance of our golden model.

\begin{table}
\caption{Synthesis results of NOPE for a 64 BS antenna, 16 UE system and comparison to existing massive MU-MIMO data detectors.}\label{tbl:comparison}
\vspace{-0.2cm}
\begin{center}
\scalebox{0.815}{
\renewcommand{\arraystretch}{1.1}      
\begin{tabular}{@{}l c c c c c @{}} 
\toprule
\multirow{2}{4em}{} & This work & Prabhu \cite{PRLE2017}         & Tang  \cite{TCZ2016}     & Peng \cite{PLZYW2017} & \!\! Casta\~neda \cite{CGS2016} \\
\midrule
System ($B\times U$) & $64\times16$ & $128\times8$ & $128\times32$ & $128\times8$ & $128\times8$ \\ 
Algorithm & {NOPE} & \!\!MMSE/ZF$^a$\!\! & MPD & MMSE & SDR \\ 
Parameters & none & $\EX$,$\No$$^c$ & $\EX$,$\No$ & $\EX$,$\No$ & $\EX$,$\No$  \\ 
Modulation & 256-QAM & 256-QAM & 256-QAM & 64-QAM & QPSK\\ 
Preproc. included & no & yes & no & yes & no \\ 
Preproc.\ quantities\!\!\!\!  & col. gains & -- & Gram mat. & -- & Gram mat. \\ 
\midrule
Results & synthesis & ASIC & ASIC & ASIC & \!\! post-layout\!\!  \\ 
Technology [nm] & 28 & 28 & 40 & 65 & 45 \\ 

Area [$\text{mm}^2$] & 0.28 & 1.10  & 0.58 & 2.57 & 0.48 \\ 
Frequency [MHz] & 800 & 300 & 425 & 680 & 560 \\ 
Throughput [Gb/s]\!\!\!\!  & 0.92 & 0.30  & 2.76 & 1.02 & 0.13 \\ 
Eff.$^b$ [Gb/s/$\text{mm}^2$]\!\! \!\! & 3.29 & 0.27  & 13.87 & 4.96 & 1.08\\ 
\bottomrule
\end{tabular}} 
\end{center}
\footnotesize $^a$this design also supports precoding; $^b$standard technology scaling rules apply; $^c$the ZF mode does not require any parameters.
\label{tbl:values}
\end{table}

\subsection{Implementation Results and Conclusion}

\fref{tbl:values} shows synthesis results for NOPE in a 28\,nm CMOS technology and compares our design to existing data detectors for massive MU-MIMO. 
We note that the numbers reported in \fref{tbl:values} for NOPE are based on synthesis results; an ASIC design is part of ongoing work.
While our design is comparable to other designs in terms of hardware efficiency, we emphasize that NOPE is completely parameter-free (other than knowledge of $\bH$ and $\bmy$), which makes it more resilient to parameter mismatch and dynamic variations of the system compared to all the other methods. 
In addition, NOPE requires a minimal amount of preprocessing, i.e., $\hat{\bmd}^2$ and $\hat{\bmd}^{-2}$, in contrast to, e.g., the design of~\cite{TCZ2016} that requires computation of the Gram which often dominates the complexity of massive MU-MIMO data detectors \cite{WYWDCS2014}.
In summary, NOPE is a robust ``fire-and-forget'' equalization algorithm for MU-MIMO systems that achieves L-MMSE performance at competitive implementation complexity.


\bibliographystyle{IEEEtran}
\bibliography{VIPabbrv,publishers,confs-jrnls,VIP_170121_NOPE,VIP}


\end{document}